\documentclass[aps,pre,amsmath,amssymb,twocolumn,groupedaddress,superscriptaddress,showpacs]{revtex4-1}

\usepackage{graphicx}
\usepackage{dcolumn}
\usepackage{bm}
\usepackage{lipsum}
\usepackage{float}
\usepackage{subfigure}
\usepackage{color}
\definecolor{ultramarine}{RGB}{0,32,96}
\definecolor{orange}{rgb}{1,0.5,0}
\definecolor{applegreen}{rgb}{0.55, 0.71, 0.0}
\definecolor{jonquil}{rgb}{0.98, 0.85, 0.37}
\definecolor{antiquebrass}{rgb}{0.8, 0.58, 0.46}

\begin{document}

\title{Thermal gas rectification using a sawtooth channel}

\author{S Sol\'orzano}
\affiliation{%
Computational Physics for Engineering Materials, Institut f. Baustoffe (IfB), ETH Zurich,
Wolfgang-Pauli-Street 27, 8093 Zurich, Switzerland
}
\author{N.A.M. Ara\'ujo} 
  \email{nmaraujo@fc.ul.pt}
  \affiliation{Departamento de F\'{\i}sica, Faculdade de Ci\^{e}ncias, Universidade de Lisboa, P-1749-016 Lisboa, Portugal, 
and Centro de F\'isica Te\'orica e Computacional, Universidade de Lisboa, P-1749-016 Lisboa, Portugal}
\author{H.J. Herrmann} 
\affiliation{%
Computational Physics for Engineering Materials, Institut f. Baustoffe (IfB), ETH Zurich,
Wolfgang-Pauli-Street 27, 8093 Zurich, Switzerland
}
\affiliation{%
Departamento de F\'isica, Universidade Federal do Cear\'a, 60451-970 Fortaleza,Cear\'a, Brazil
}

\begin{abstract}
We study the rectification of a two-dimensional thermal gas in a channel of
asymmetric dissipative walls. For an ensemble of smooth Lennard-Jones
particles, our numerical simulations reveal a non-monotonic dependence of the
flux on the thermostat temperature, channel asymmetry, and particle density,
with three distinct regimes. Theoretical arguments are developed to shed light
on the functional dependence of the flux on the model parameters.
\end{abstract}
 
\maketitle

\section{\label{introduction}Introduction}
Systems where fluctuations are rectified into directed motion are known as
Brownian motors or ratchet devices~\cite{RevModPhys.81.387,Reimann200257}.
According to the second law of thermodynamics, these systems ought to be
impossible in equilibrium~\cite{Feynman1963_3,Reimann200257}. This is why
broken spatial symmetry and non equilibrium conditions are a key feature for
their operation~\cite{mdemon,ANDP:ANDP200410121}. Brownian motors are relevant
in a number of situations from biological processes to devices for particle
segregation~\cite{PhysRevE.58.7781,PhysRevE.61.312,matthiasasymmetric2003,rousselet_directional_1994}
and
transport~\cite{PhysRevLett.88.168301,PhysRevLett.74.1504,PhysRevLett.72.2652}.
For instance, asymmetric objects (e.g., wedge shapes) immersed in a granular
gas tend to move~\cite{PhysRevE.75.061124,0295-5075-77-5-50003} or
rotate~\cite{PhysRevLett.93.090601,PhysRevLett.104.248001,PhysRevLett.110.120601,PhysRevE.83.031310}
in a preferential direction, provided that particle/object collisions are
dissipative. Also, granular particles enclosed in a vibrating sawtooth-shaped
channel flow along a preferential direction defined by the asymmetry of the
channel~\cite{PhysRevE.88.042201}. Similar results are also observed for
microscopic particles in an asymmetric channel under the action of a pulsating
potential~\cite{PhysRevE.61.312} .
There are also various examples of active matter systems ranging from bacteria at microscopic
scale~\cite{PhysRevLett.102.048104,DiLeonardo25052010,Sokolov19012010}, to
centimeter scale bots~\cite{0295-5075-102-5-50007}, up to
pedestrians~\cite{PhysRevX.6.011003} where rectification can be induced by
spatial asymmetries. 

Although these systems rectify the motion of the particles or objects immersed within the particle bath, they still require either a pulsating potential or active particles. Further examples, that in some sense relax these requirements, include rectification using differentiated noise sources~\cite{holubec2017thermal,ryabov2016transport} or asymmetric piston models ~\cite{PhysRevE.87.040101,0295-5075-82-5-50008,PhysRevE.85.021310} that show rectification effects, even when working at a single temperature, provided there is friction and the particle-piston collisions are different on both sides of the piston. So far there are only few examples~\cite{PhysRevE.88.052124,0295-5075-97-2-20001} in which the motion of a single particle in a single dimension is rectified without external driving forces. In the present work, we provide a novel example of collective particle motion rectification in two dimensions. We show that the motion of a gas of Lennard-Jones particles can be rectified without external pulsating potentials only by means of dissipation and broken spatial asymmetry. To this end, we consider a two-dimensional gas of particles in a fixed (not moving) asymmetric sawtooth-shaped channel and study how the overall flux depends on the different model parameters. We expect that this idea can find applications in fields such as microfluidic or lab on a chip set ups.

The paper is organized as follows. In Section~\ref{system}, the model and methods are introduced. The results are discussed in Section~\ref{results} and we draw some conclusions in Section~\ref{conclusion}.

\section{\label{system}Model and methods}  
\begin{figure}[t]
\centering
\includegraphics[width=\columnwidth]{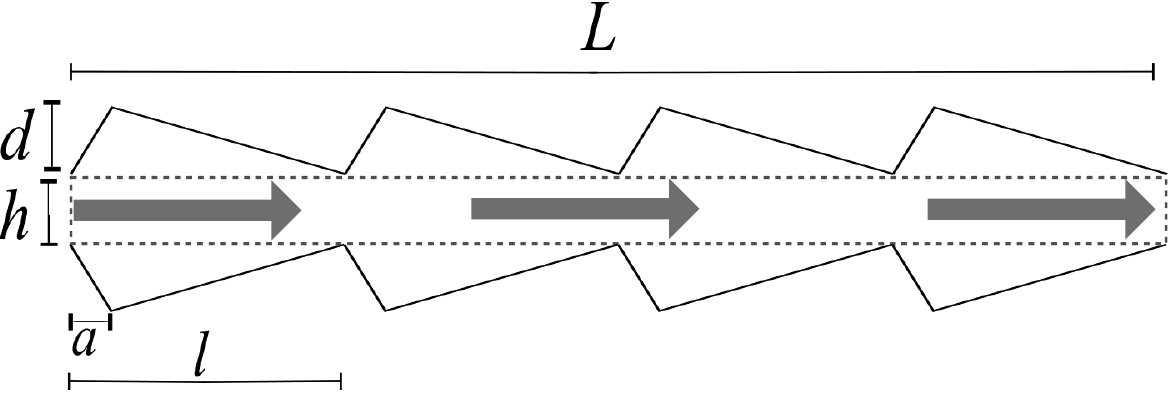}
\caption{Schematic representation of the channel of size $L$, with $N=4$
	cells. The shape of each cell is characterized by four lengths: the
	linear length $l$, the aperture size $h$, and the horizontal position
	$a$ and height $d$ of the peak. The depicted channel is classified as right asymmetric (see text) and the arrows indicate the corresponding direction of particle flow. The dashed lines delimit the region where particles are initially released.\label{fig::fig1}}
\end{figure}
We consider a two-dimensional sawtooth channel of linear length $L$ consisting
of a sequence of $N$ equal cells, as represented in Fig.~\ref{fig::fig1}, with
periodic boundary conditions along the horizontal axis. The geometry of the
channel is characterized by four lengths: the length of each cell $l=L/N$, the
aperture size $h$, and the horizontal position $a$ and height $d$ of the edge.
To systematically study the dependence on the asymmetry of the channel, we fix
$h$ and define the adimensional asymmetry coefficient $\alpha$ as,
\begin{equation}\label{eq::alpha}
\alpha= 1-2\frac{a}{l},
\end{equation}
where $\alpha\in [-1,1]$. For $\alpha=0$ the channel is symmetric with respect
to the vertical axis, while for $\alpha=\pm1$ the cells look triangular.  We
classify channels of negative and positive $\alpha$ as left and right
asymmetric, respectively.

\begin{figure*}[t]
\centering
\includegraphics[width=\textwidth]{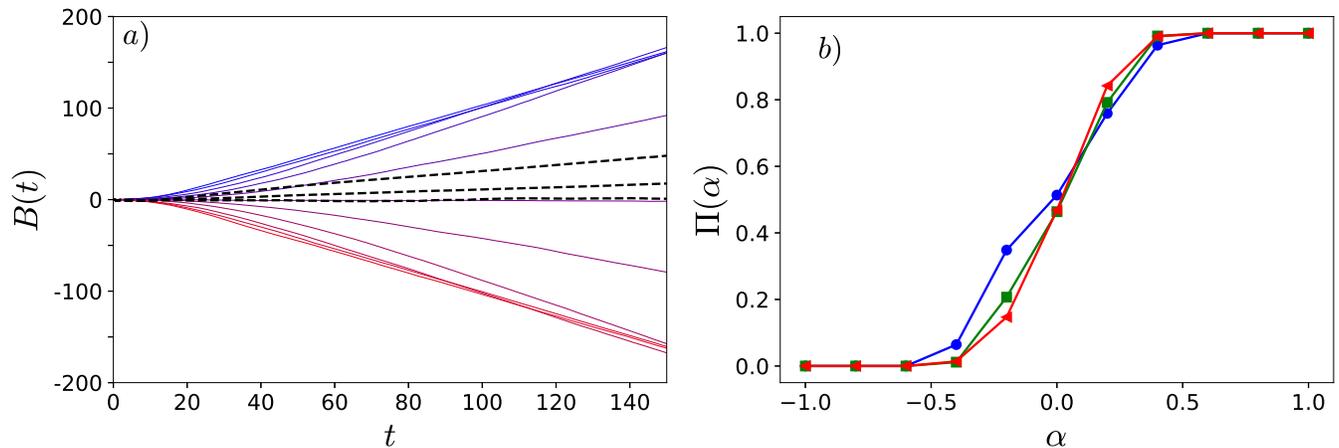}
\caption{(Color online) a) Time evolution of the balance $B(t)$ for different
	values of $\alpha=\{-1.0,-0.8,-0.6,-0.4,-0.2,0,0.2,0.4,0.6,0.8,1.0\}$
	(from bottom to top). Results are averages over $500$ samples of
	systems of $100$ particles at a thermostat temperature $T=2.5$ and
	$\gamma=1$.The black dashed lines correspond to simulations with $\alpha=\{0, 0.8, 1\}$ and $d_c=5\sigma$ which show that the rectification phenomena is not unique to the choice $d_c=2.5\sigma$  b) Fraction of samples $\Pi$ for which $B(t)>0$, for $T=2.5$, $\rho=0.143$, and $N=250$ (blue circles), $N=500$ (green squares), $N=1000$ (red triangles) .} \label{fig::fig2}
\end{figure*}
\begin{figure*}[t]
 \centering
 \includegraphics[width=\textwidth]{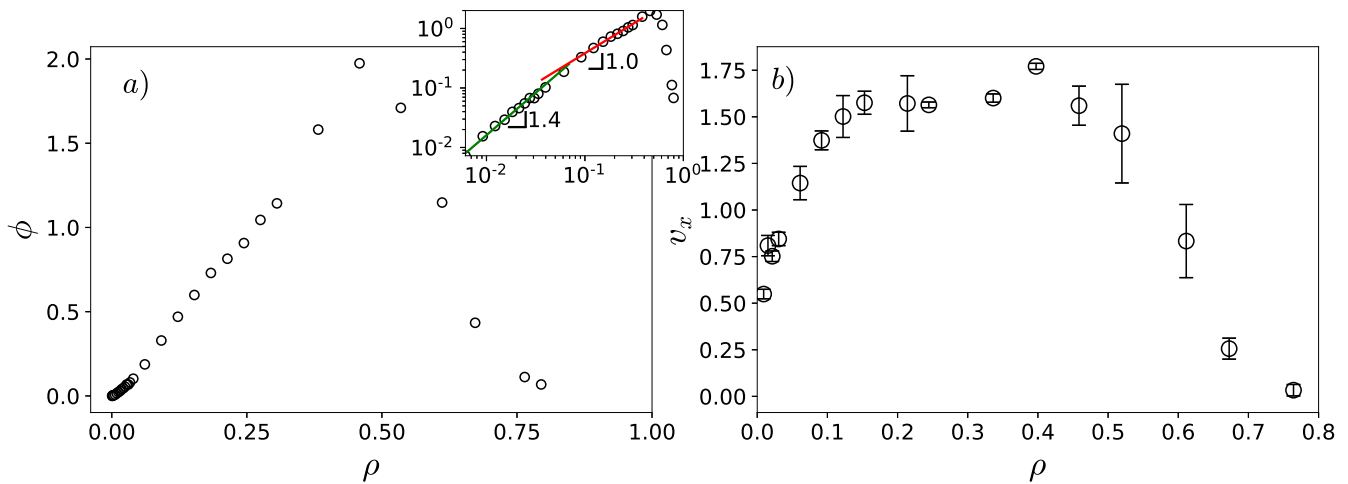}
 \caption{(Color online) a) Flux $\phi$ as a function of the density $\rho$,
	 for $T=2.5$ and $\alpha=1$.  The two initial regimes are shown in the
	 inset in a double logarithmic plot, where the dashed lines represent
	 the power-law fits, $\phi\sim\rho^\beta$, with $\beta=1.47\pm0.07$
	 and $\beta=1.01\pm0.04$ for the low and intermediate density regimes,
	 respectively. b) Average horizontal component of the particle
	 velocity ($v_x$) as a function of the density, for the same set of
	 parameters. \label{fig::fig3}}
\end{figure*}

We consider a gas of particles interacting pairwise, where the force of
particle $j$ on particle $i$ is conservative and given by
$\mathbf{F}_{ij}=-\mathbf{\nabla}_iU_\mathrm{LJ}$. $U_\mathrm{LJ}$ is the
$12$-$6$-Lennard-Jones potential,
\begin{equation}\label{eq::LJPotential}
U_{\text{LJ}}\left(r_{ij}\right) =
4\epsilon\left[\left(\frac{\sigma}{r_{ij}}\right)^{12} -
\left(\frac{\sigma}{r_{ij}}\right)^{6}\right] \ \ ,
\end{equation}
where $r_{ij}=\left|\mathbf{r}_{j}-\mathbf{r}_{i}\right|$, and $\mathbf{r}_i$ and
$\mathbf{r}_j$ are the positions of particles $i$ and $j$, respectively. 
$\epsilon$ corresponds to depth of the potential well which is located at $r_{m}=2^{1/6}\sigma$. The force $\mathbf{F}_\mathrm{iw}$ of the wall on a particle $i$ is described as
the superposition of two contributions: a conservative force,
$\mathbf{F}_\mathrm{iw}^c$, and a dissipative one, $\mathbf{F}_\mathrm{iw}^d$.
The conservative force is described as a Lennard-Jones interaction with the
closest point on the wall, with the same $\epsilon$ and $\sigma$ of the
particle/particle interaction. The dissipative force is given by,

\begin{equation}\label{eq::disswall}
\mathbf{F}_\mathrm{iw}^d=-\gamma\left(\dot{\mathbf{r}}_{i}\cdot\mathbf{\hat{n}}_{iw}\right)\mathbf{\hat{n}}_{iw}
\ \ ,
\end{equation}
where
${\mathbf{\hat{n}}_{iw}}=\frac{{\mathbf{r}_{i}}-{\mathbf{r}_{w}}}{r_{iw}}$ is
the unit vector pointing from the closest point on the wall $\mathbf{r}_{w}$
to the particle $i$ and $\gamma\geq0$ is a friction constant. The
particle/wall interaction is conservative for $\gamma=0$ and dissipative
otherwise. Interactions with the wall are truncated at a 
cutoff distance $d_c=2.5\sigma$ and if the particle is within the cutoff distance 
of multiple points or walls the contributions are superimposed. The particle/wall interaction 
model was chosen to study the effect of the wall geometry on the particles dynamics. 
It is assumed that the particles locally bounce off the wall thus the use of a cutoff 
distance and the nearest point prescription. $d_c=2.5\sigma$ has been found to be 
a reasonable cut-off Ref.\cite{csl}. The analysis of more complex particle/wall
interaction models where the walls are directly modelled as a fixed set 
of particles is left for future work.

We performed canonical molecular dynamics simulations, using the Nos\'e-Hoover
thermostat~\cite{csl,granular,PhysRevA.31.1695}. Accordingly, the equation of
motion of particle $i$ is,
\begin{align}
	&\ddot{\mathbf{r}}_i=\frac{1}{m_i}\left(\sum_{j\neq
	i}\mathbf{F}_{ij}(\mathbf{r}_{ij}) + \mathbf{F}_\mathrm{iw}\right) +
	\mathbf{F}_\mathrm{NH}\ \ ,\label{eq::motion} \\
	&\mathbf{F}_\mathrm{NH} = -\xi\dot{\mathbf{x}}\ \ ,\nonumber
\end{align}
where $\mathbf{r}_{ij}=\mathbf{r}_j-\mathbf{r}_i$, $m_i$ is the particle mass 
and $\mathbf{F}_\mathrm{NH}$ is the force per unit of mass, resulting from 
the coupling with the thermostat~\cite{PhysRevA.31.1695}. $\xi$ is the variable that describes the 
thermostat and its dynamics is given by 
\begin{equation}
\dot{\xi}=\frac{1}{Q}\left(\sum_{i=1}^{N}\frac{\mathbf{p}^{2}_{i}}{m_{i}}-(3N+1)k_{B}T\right)\label{eq::motionb},
\end{equation}
where $Q$ is a parameter known as ``thermal inertia'' that throughout the work was set to $Q=0.05$, $k_{B}$ is the Boltzmann constant, $N$ is the number of particles and $T$ is the thermostat temperature.

For simplicity, we set $m_i\equiv m$ and consider reduced units, such that: mass is in units of $m$, distance in
units of $\sigma$, and energy in units of $\epsilon$.  The equations of motion
are integrated using a fifth-order predictor-corrector algorithm, with a time
step $dt=10^{-5}$ and we run the simulation up to $t=165$.

To generate the initial configurations, all particles were released within the
region delimited by the dashed lines in Fig.~\ref{fig::fig1}, with an initial
velocity drawn from a uniform distribution of zero mean.  Particles are
thermalized at the thermostat temperature within the dashed region,
considering periodic boundary conditions along the horizontal direction and
reflective top and bottom boundaries, without interacting with the channel
walls. At $t=15$, the constraint imposed by the dashed lines is removed and
particles move inside the channel, following the dynamics described by
Eq.~(\ref{eq::motion}).

\section{\label{results}Results}

\begin{figure}[t]
 \centering
\includegraphics[width=\columnwidth]{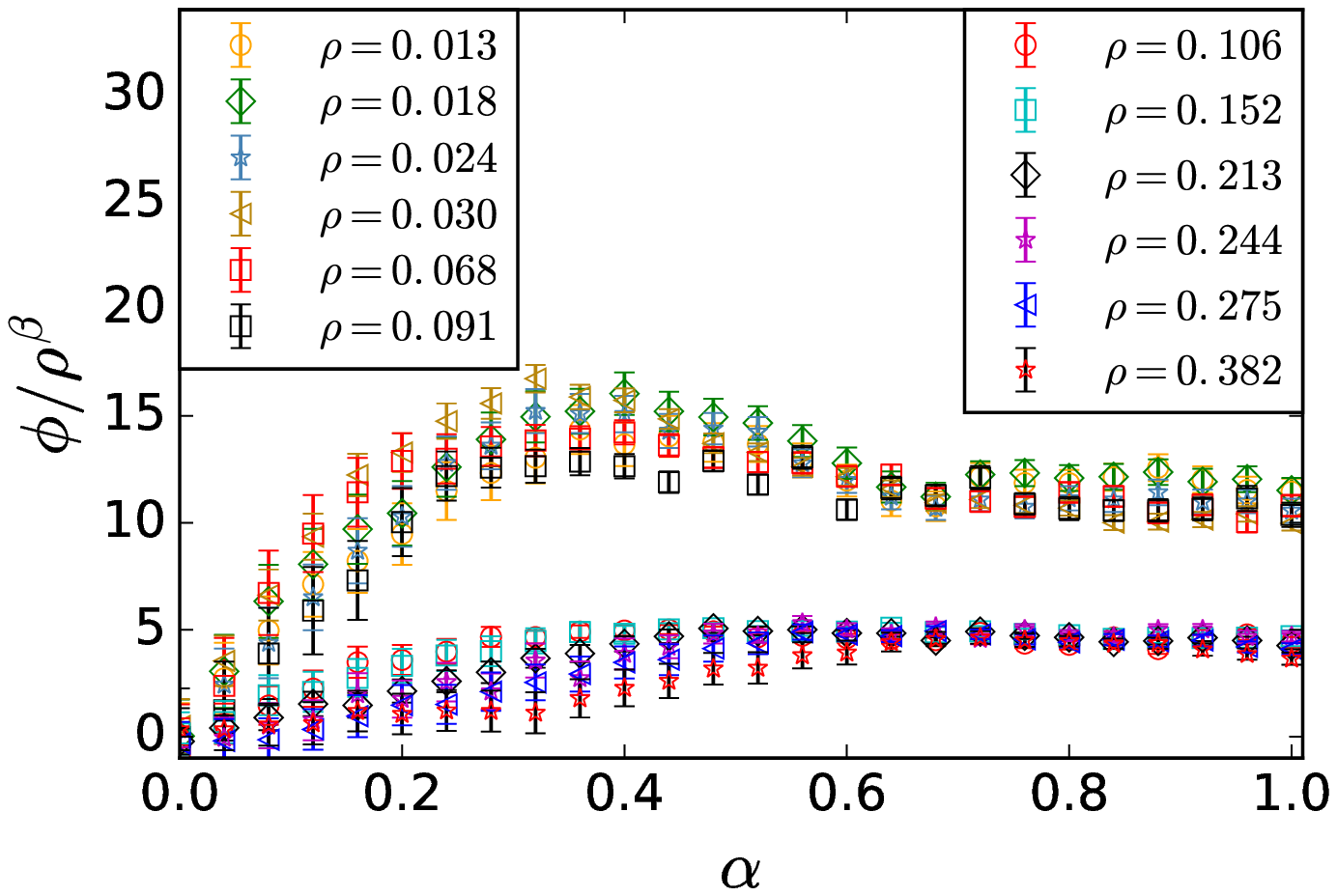}
 \caption{(Color online) a) Flux $\phi$ rescaled by $\rho^\beta$ as a function of
	 $\alpha$ for $T=2.5$ and different values of the density. The values reported on the left(right) legend
	 were rescaled using $\beta=1.47$($\beta=1.01$).\label{fig::fig4}}
\end{figure}

\begin{figure}[t]
\centering
\includegraphics[width=\columnwidth]{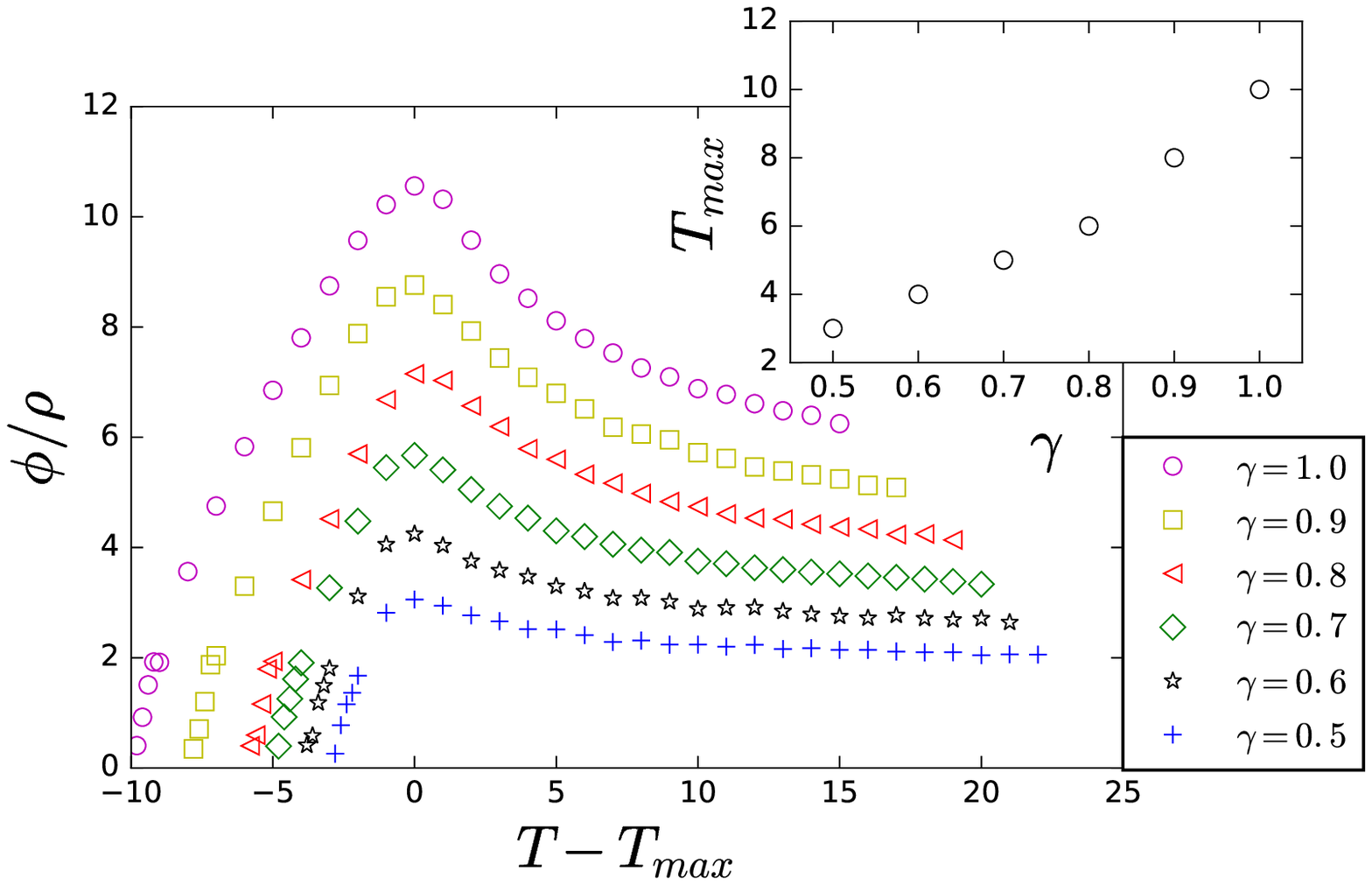}
\caption{(Color online) Scaled flux, $\phi/\rho$, as a function of the
	thermostat temperature, for $\alpha=1$, $\rho=0.0917$, and different
	values of the friction constant, $\gamma$. The thermostat temperature
	was shifted by $T_\mathrm{max}$, defined as the optimal temperature at
which a maximum is observed in the flux. The dependence of $T_\mathrm{max}$ on
$\gamma$ is shown in the inset. \label{fig::fig5} }
\end{figure}

To characterize the effect of the asymmetry of the channel walls on the overall
flux, we fixed $N=24$, $l=15$, $d=3$, and $h=6$, and performed simulations for
different values of the asymmetry coefficient $\alpha\in [-1,1]$. To reduce
statistical noise, instead of directly measuring the outlet flux $\phi(t)$, we
introduce an integrated quantity $B(t)$, which we call balance. $B(t)$ is defined
as the difference between the cumulative number of particles crossing the
rightmost boundary from the left to the right and the ones crossing it in the
opposite direction, up to time $t$. In the continuum limit,
\begin{equation}\label{eq::balance-flux}
\phi(t)=\frac{1}{h}\dot{B}(t) \ \ .
\end{equation}
Asymptotically, we expect that the balance scales linearly in time, and so
we estimated the flux from a linear regression fit of the curve $B(t)$ in the
linear regime.

Figure~\ref{fig::fig2}(a) shows the balance as a function of time for
different values of $\alpha$. Clearly, spontaneous flow emerges as a result of
the asymmetry of the channel walls. For right symmetric channels ($\alpha>0$),
the flow is from the left to the right, while for left symmetric channels
($\alpha<0$), the flow is in the opposite direction. Figure~\ref{fig::fig2}(a) also shows examples for $\alpha=\{0,0.8,1\}$, for a particle-wall cutoff distance of $5\sigma$. These examples show that the observed rectification of the particle motion is still observed for a larger cutoff distance, but the quantitative values of the flux are obviously different.

To analyze the transition
at $\alpha=0$, we define $\Pi$ as the fraction of samples where $B(t)>0$ for
large values of $t$ ($t=165$). The dependence of $\Pi$ on $\alpha$ is in
Fig.~\ref{fig::fig2}(b), for three different sizes of the channel. One sees that $\Pi$ is
$0.5$ for $\alpha=0$ and it seems to converge to a step function as the system size
increases.

The dependence of the flux on the density is shown in Fig.~\ref{fig::fig3}(a),
for $\alpha=1$ and $T=2.5$.  One clearly observes an optimal density
($\rho_\mathrm{opt}\approx0.45$) at which the flux is maximized. The data for
$\rho<\rho_\mathrm{opt}$ suggests two different regimes (see inset of
Fig.~\ref{fig::fig3}(a)): a low-density regime, for $\rho<0.1$, and an
intermediate-density regime, for $0.1<\rho<\rho_\mathrm{opt}$. It is expected  
that the flux of particles is a monotonically increasing function of the density up to the point where it 
either saturates or start to decrease. For systems where the particle/particle collisions are more frequent
than the particle/wall collisions, kinetic theory~\cite{cercignani1988boltzmann} suggests that the flux is 
linear in the density. However, for low densities, most of the particle collisions are with the wall; This
implies that the flux is mostly determined by the chance of a particle to bounce off the wall and eventually
cross the boundary at either end of the channel thus increasing or decreasing the net flux. In this case the
monotonic increase of the flux is not necessarily linear. The numerical data suggest a power law 
scaling: $\phi\sim\rho^\beta$. Assuming a power-law scaling, we estimate $\beta=1.47\pm0.07$, for
the low-density regime, and $\beta=1.01\pm0.04$, for the intermediate-density one.

Figure~\ref{fig::fig3}(b) shows the average horizontal component of the
velocity ($v_x$) as a function of the density. For the first regime, $v_x$
increases with the density thus, the enhancement of the flux with the density
stems from an increase in the number particles per cell and possible collective
effects affecting the particle velocity. Similar flux enhancement was reported in the context of comb systems, where the comb tooth would take the role of the sawtooth~\cite{PhysRevLett.115.220601}. However, in Ref.~\cite{PhysRevLett.115.220601}, the flux enhancement is observed for high densities and is related to the saturation of the traps. Here, instead, we observe a flux increase for much lower densities suggesting a different mechanism. 

By contrast, for the intermediate-density regime, $v_x$ does not significantly change with the density and so the flux only increases due to an increase in the number of particles per cell, yielding a linear scaling. For $\rho>\rho_\mathrm{opt}$, a third regime is observed, for which the flux simply decreases with the density
due to crowding effects. That is, as the density increases, the available space 
for a particle to move diminishes. Thus, the motion of a single particle is strongly constrained by the presence of others. This in turn implies that for a particle to move over an extended region, it requires a concerted re-arrangement of several other particles. As the density increases, the chances that this concerted motion leads to a majority of particles moving in a preferred direction diminishes. Instead, it is more likely that the particles re-arrange by moving with no preferred direction leading to a decrease in the flux. Notice that this is consistent with the fact that $v_x$ approaches zero (Fig.~\ref{fig::fig3}(b)) which implies that roughly the same number of particles travel in each direction.

To study the dependence on $\alpha$, we plot in Fig.~\ref{fig::fig4}, the flux
rescaled by $\rho^\beta$, using the estimated values of $\beta$ for the
corresponding regime. We observe a data collapse for each regime, suggesting 
that the power-law scaling is resilient over the entire range 
of α values. The low density regime shows an optimal value of α whereas the intermediate regime shows instead a nearly constant be- havior. We think that, for high-enough density, the rate of particle-particle collisions is significantly higher than the one of particle-wall collisions and thus the geometry of the walls does not play a significant role on the overall dynamics. By contrast, for low density, the rates of particle-particle and particle-wall collisions are comparable and a competition between the two is observed, leading to the maximum in the flux. Although there is some dispersion on the data collapse, it is clear that there are two distinct regimes. The origin of the dispersion can be due to finite-size effects or scaling corrections. The study of the nature of the transition and crossover between these two regimes requires further study, that is beyond the scope of this work. It is interesting to notice that the results for low densities in Fig.~\ref{fig::fig4} are similar to those obtained by A. Sarracino~\cite{PhysRevE.88.052124}. Namely, the quantity that indicates the presence of motion rectification ($\phi$ in our case and $\langle V\rangle$ in Ref.~\cite{PhysRevE.88.052124}) shows a qualitatively similar non-monotonic behavior as a function of the asymmetry parameter. In both cases, it vanishes for the symmetric case and initially grows with the asymmetry, having a maximum for intermediate values, and then decreasing towards a saturation value for large values of the asymmetry parameter.

The dependence of the flux on the thermostat temperature is shown in
Fig.~\ref{fig::fig5}, for different values of $\gamma$. For the entire range
of values of $\gamma$, a maximum is observed at an optimal temperature,
$T_\mathrm{max}$, that increases with $\gamma$ (see
inset). Also, the optimal flux grows with dissipation (increasing
$\gamma$). Note that, for $\gamma=0$ the overall flux vanishes and thus a
dissipative interaction with the walls is necessary to rectify the thermal
motion of gas. This is consistent with the work of Prost \textit{et al.} that
suggests that time-reversal symmetry of trajectories needs to be broken to
obtain rectification from asymmetric walls~\cite{Prost94}.

Finally, to quantify the flux for a specific system, let us consider a channel
of total length $L=216\,\mu\text{m}$, single cell length $l=9\,\mu\text{m}$,
$\alpha=0$, $p=1.8\,\mu\text{m}$ and $h=3.7\,\mu\text{m}$ at room temperature,
with colloidal particles of $\sigma=6\times 10^{-7}\,\text{m}$, $m=2.49\times
10^{-16}\,\text{kg}$, and $\gamma=2\times
10^{-14}\,\text{kg}/\text{s}$~\cite{doi:10.1021/jp407247y,colloid2}. If we
assume, $\epsilon/k_{b}=0.01414\,\text{K}$ (where $k_{b}$ is the Boltzmann
constant), we obtain a flow velocity $\phi/\rho\approx
42\,\mu\text{m}/\text{s}$.

\section{\label{conclusion} Conclusion}
In this work, we systematically study the dependence of the rectification of
the motion of a thermal gas on a channel of asymmetric dissipative walls. We
found that the overall flux enhances with the friction constant of the
particle/wall interaction and that it shows a nonmonotonic dependence on three
other model parameters, namely, the thermostat temperature, channel asymmetry,
and particle density. For the dependence of the flux on the density of
particles, we found three different regimes. For low density, the flux scales
superlinearly with the density, as collective effects lead also to an increase
in the horizontal component of particle velocity. For intermediate density, the
horizontal component of particle velocity saturates at a constant value and the
overall flux scales linearly with the density. Finally, above an optimal value
of the density, the flux monotonically decreases due to crowding effects. Future
work might consider different geometries and a generalization to the
three-dimensional case. The effect of different dissipation mechanisms as well
as particle shapes are still open questions.

\acknowledgments{We acknowledge financial support from the Brazilian institute
	INCT-SC and grant number FP7-319968 of the European Research Council.
	NA acknowledges financial support from the Portuguese Foundation for
	Science and Technology (FCT) under Contract no. UID/FIS/00618/2013 and
	from the Luso-American Development Foundation (FLAD), FLAD/NSF, Proj.
273/2016.}

\bibliography{bibliography}
\end{document}